\documentclass{elsart}
\usepackage{epsfig}

\begin{document}

\begin{frontmatter}

\title{A Distributed Algorithm to Find \boldmath $k$-Dominating Sets}

\author[ldpvcb]{Lucia D. Penso\thanksref{ldp}}
\ead{lucia@cs.brown.edu}
\author[ldpvcb]{Valmir C. Barbosa\corauthref{vcb}}
\ead{valmir@cos.ufrj.br}
\thanks[ldp]{Currently at the Computer Science Department, Brown University,
Providence, RI 02912, USA.}
\corauth[vcb]{Corresponding author.}

\address[ldpvcb]
{Programa de Engenharia de Sistemas e Computa\c c\~ao, COPPE,\\ 
Universidade Federal do Rio de Janeiro,\\
Caixa Postal 68511, 21941-972 Rio de Janeiro - RJ, Brazil}

\begin{abstract}
We consider a connected undirected graph $G(n,m)$ with $n$ nodes and $m$ edges.
A $k$-dominating set $D$ in $G$ is a set of nodes having the property that every
node in $G$ is at most $k$ edges away from at least one node in $D$. Finding a
$k$-dominating set of minimum size is NP-hard. We give a new synchronous
distributed algorithm to find a $k$-dominating set in $G$ of size no greater
than $\left\lfloor n/(k+1)\right\rfloor$. Our algorithm requires $O(k\log^*n)$
time and $O(m\log k+n\log k\log^*n)$ messages to run. It has the same time
complexity as the best currently known algorithm, but improves on that
algorithm's  message complexity and is, in addition, conceptually simpler.
\end{abstract}

\begin{keyword}
$k$-dominating sets \sep distributed algorithms \sep graph algorithms
\end{keyword}

\end{frontmatter}

\section{introduction}\label{intr}

Let $G(n,m)$ be a connected undirected graph with $n$ nodes and $m$ edges. For
$k\le n-1$, a {\em $k$-dominating set\/} $D$ in $G$ is a set of nodes with the
property that every node in $G$ is at most $k$ edges away from at least one of
the nodes of $D$. The problem of finding $k$-dominating sets of relatively small
sizes is important in a variety of contexts, including multicast systems
\cite{wz01}, the placement of servers in a computer network \cite{bkp93}, the
caching of replicas in database and operating systems \cite{p90}, and message
routing with sparse tables \cite{pu89}.

Finding a $k$-dominating set in $G$ with the least possible number of nodes is
an NP-hard problem \cite{gj79}, so one normally settles for a set of small size
that is not necessarily optimal. In general, the small size to be sought is at
most $\left\lfloor n/(k+1)\right\rfloor$, since it can be argued that a
$k$-dominating set with no more than this number of nodes always exists
\cite{kp98}, and likewise that a connected graph on $n$ nodes necessarily exists
for which every $k$-dominating set has at least
$\left\lfloor n/(k+1)\right\rfloor$ nodes \cite{p99}.

The argument for $\left\lfloor n/(k+1)\right\rfloor$ as an upper bound is
instructive in the present context, and goes as follows \cite{kp98}. Let $T$ be
a rooted spanning tree of $G$ and $D_1,\ldots,D_{k+1}$ a partition of its nodes
such that, for $0\le\ell\le k$, every node in $D_{\ell+1}$ is away from the root
a number $x_\ell$ of tree edges such that $x_\ell\bmod(k+1)=\ell$. This
partition can be constructed easily by traversing $T$ breadth-first from the
root and assigning every new layer of nodes circularly to the sets
$D_1,\ldots,D_{k+1}$. Clearly, every one of these sets is a $k$-dominating set
in $G$. Also, because they partition the graph's node set, and considering that
$n\ge k+1$, it must be that at least one of them has no more than
$\left\lfloor n/(k+1)\right\rfloor$ nodes.

Our topic in this paper is finding a $k$-dominating set in $G$ having no more
than $\left\lfloor n/(k+1)\right\rfloor$ nodes by means of a synchronous
distributed computation on $G$. The model of distributed computation that we
adopt is the standard fully synchronous model \cite{b96}. In this model, the
nodes of $G$ are processors that function in lockstep at the occurrence of clock
pulses, and its edges are bidirectional communication channels that deliver
messages between their end nodes before the clock pulse that follows the sending
of the message occurs. Time is measured by counting clock pulses.

The current best synchronous algorithm to find a $k$-dominating set in $G$ is
from \cite{kp98}, and is henceforth referred to as Algorithm KP. It proceeds in
two stages: the first stage partitions $G$ into the trees of a rooted spanning
forest $F$, each having at least $k+1$ nodes and height $O(k)$; the second stage
approaches each tree $U\in F$ as described earlier for the spanning tree $T$ and
partitions its nodes into the sets $D^U_1,\ldots,D^U_{k+1}$. If $f$ is the
number of trees in $F$, then the $k$-dominating set output by the algorithm is
$D=D_{\ell_1}^{U_1}\cup\cdots\cup D_{\ell_f}^{U_f}$, where $U_1,\ldots,U_f$ are
the trees of $F$ and $D_{\ell_i}^{U_i}$ is the smallest set of
$D_1^{U_i},\ldots,D_{k+1}^{U_i}$ for $1\le i\le f$. If $n_U$ is the number of
nodes of $U\in F$, then
\begin{eqnarray*}
\vert D\vert
&=&\vert D_{\ell_1}^{U_1}\vert+\cdots+\vert D_{\ell_f}^{U_f}\vert\\
&\le&\sum_{U\in F}\left\lfloor\frac{n_U}{k+1}\right\rfloor\\
&\le&\left\lfloor\frac{\sum_{U\in F}n_U}{k+1}\right\rfloor\\
&=&\left\lfloor\frac{n}{k+1}\right\rfloor,
\end{eqnarray*}
since $n_U\ge k+1$ for all $U\in F$.

While the second stage of Algorithm KP can be easily implemented within the
bounds of $O(k)$ time and $O(n)$ messages, its first stage is based on an
arcane combination of previously developed algorithms for related problems
\cite{gkp98,gps87,ps92}, resulting in a time complexity of $O(k\log^*n)$ and a
message complexity of $O(m\log k+n^2\log^*n)$. The latter, incidentally, is our
best estimate of what is really involved, in terms of communication needs, in
Algorithm KP---such needs are thoroughly ignored in \cite{kp98}, but this
message complexity seems to follow from the message complexities of the
algorithm's building blocks.

In this paper, we introduce a new synchronous distributed algorithm for finding
a $k$-dominating set of no more than $\left\lfloor n/(k+1)\right\rfloor$ nodes
in $G$. Like Algorithm KP, our algorithm too comprises two subsequent stages,
each having the same goal as its counterpart in Algorithm KP. The second stage,
in particular, is exactly the same as Algorithm KP's.

Our contribution is the introduction of a new algorithm for the partition of
$G$ into the trees of $F$. When compared to Algorithm KP, our algorithm has the
same complexity of $O(k\log^*n)$ time while improving on the message complexity,
which in our case is of $O(m\log k+n\log k\log^*n)$. We also find our algorithm
to be conceptually simpler than Algorithm KP, which can probably be attributed
to the fact that it was designed from scratch with the partitioning problem in
mind. While our algorithm simply generates a sequence of ``meta-graphs,'' the
last of which has nodes that directly give the rooted trees of $F$, Algorithm KP
reduces the partition problem to other related problems and then combines
algorithms for those problems into building a solution to the partition problem.
Henceforth, we let the algorithm we introduce be called Algorithm PB.

The following is how the remainder of the paper is organized. The first stage
of Algorithm PB is introduced in Section \ref{alg} and analyzed for correctness
and complexity in Section \ref{corr}. Concluding remarks are given in Section
\ref{concl}.

\section{The algorithm}\label{alg}

In this section we introduce the first stage of Algorithm PB. This stage finds
a rooted spanning forest $F$ in $G$, each of whose trees has at least $k+1$
nodes and $O(k)$ height, and is referred to in the sequel as Partition\_$G$.

Partition\_$G$ starts by letting the node set of $G$ be the node set of a
directed graph $\vec G_0$, and proceeds from there in
$\left\lceil\log(k+1)\right\rceil$ phases. For
$0\le i\le\left\lceil\log(k+1)\right\rceil-1$, phase $i$ first builds the edge
set of $\vec G_i$ and then begins the transformation of $\vec G_i$ into another
directed graph, $\vec G_{i+1}$, by clustering the nodes of $\vec G_i$ together
to form the nodes of $\vec G_{i+1}$. Each node of $\vec G_i$ stands for a rooted
tree in $G$, and this clustering is performed in such a way that not only is
each resulting node of $\vec G_{i+1}$ also a rooted tree in $G$, but one that
has at least $2^{i+1}$ nodes and $O(2^{i+1})$ height. After completion of phase
$\left\lceil\log(k+1)\right\rceil-1$, the node set of
$\vec G_{\lceil\log(k+1)\rceil}$ represents the desired rooted spanning forest
$F$ (earlier termination is also possible, as we discuss shortly).

If $x$ and $y$ are nodes of $\vec G_i$, then we say that $x$ and $y$ are
{\em potential neighbors\/} in $\vec G_i$ if an edge exists in $G$ joining some
node in the rooted tree represented by $x$ to some node in the rooted tree
represented by $y$. We say that they are {\em neighbors\/} in $\vec G_i$ if a
directed edge exists between them. A node that has no neighbors is
{\em isolated}. If $x$ and $y$ are neighbors in $\vec G_i$, then we use $x\to y$
to indicate that the edge between $x$ and $y$ is directed from $x$ to $y$. In
this case, we say that $x$ is an {\em upstream\/} neighbor of $y$, which in turn
is an {\em downstream\/} neighbor of $x$.

Partition\_$G$ is based on manipulations of node {\em identifiers}, which we
assume to be a distinct nonnegative integer for every node in $G$. The
identifier of node $x$ in $\vec G_i$, denoted by ${\it id}(x)$, is the
identifier of the root of its tree. If a node's identifier is less than those of
all its neighbors, then we call the node a {\em local minimum}. If it is
greater, then we call it a {\em local maximum}. The following is how
Partition\_$G$ works during phase $i$. We use $\log^{(t)}n$ to denote
$\log\cdots\log n$, where $\log$ is repeated $t$ times.

{\bf Step 1.} Find the edges of $\vec G_i$:

\begin{enumerate}
\item[1a.] Let each node of $\vec G_i$ be {\em inactive}, if the height of the
corresponding rooted tree is at least $2^{i+1}$, or {\em active}, otherwise.
\item[1b.] For every active node $x$ of $\vec G_i$, find the potential neighbors
of $x$ in $\vec G_i$. If no potential neighbors are found for any node, then
halt and exit Partition\_$G$.
\item[1c.] For each active node $x$ of $\vec G_i$, let $y$ be the active
potential neighbor of $x$ with the least identifier. If $x$ has no active
potential neighbors, then let $y$ be the (inactive) potential neighbor of $x$
having the least identifier. Add $x\to y$ to the edge set of $\vec G_i$, thus
making $x$ and $y$ neighbors in $\vec G_i$.
\end{enumerate}

{\em Remark.} If no neighbors are found for any node in Step 1b, then in
reality $\vec G_i$ has one single node that encompasses all the nodes of $G$ and
therefore corresponds to a rooted spanning tree of $G$. In this case,
Partition\_$G$ terminates prematurely, that is, before completing all
$\left\lceil\log(k+1)\right\rceil$ phases.

{\em Remark.} At the end of Step 1c, every active node of $\vec G_i$ has exactly
one downstream neighbor, while every inactive node has none. Similarly, both
active nodes whose downstream neighbor is active and inactive nodes may have
between zero and some positive number of upstream neighbors. An active node
whose downstream neighbor is inactive has no upstream neighbors.

{\bf Step 2.} Find the nodes of $\vec G_{i+1}$:

\begin{enumerate}
\item[2a.] If $x\to y$ is an edge of $\vec G_i$ such that $x$ is an active node
and $y$ an inactive node, then combine $x$ into $y$ by creating a single node
whose identifier remains ${\it id}(y)$. Let $X$ be the set of active nodes of
$\vec G_i$ that are not isolated.
\item[2b.] For $x\in X$, let $Z(x)\subseteq X$ be the set of upstream neighbors
of $x$. If $Z(x)\neq\emptyset$, then let $z$ be the member of $Z(x)$ having the
least identifier. For $y\in Z(x)$ such that $y\neq z$, check whether
$Z(y)=\emptyset$. In the affirmative case, combine $y$ into $x$. Otherwise
(i.e., $Z(y)\neq\emptyset$), eliminate edge $y\to x$. Let $X$ be the set of
active nodes of $\vec G_i$ that are not isolated.
\item[2c.] For $x\in X$, if $x$ is a local minimum, then combine its (at most
two) neighbors into it and make it isolated by eliminating edges from $\vec G_i$
appropriately. Also, combine into the newly-formed node any node in $X$ that may
have become isolated. Let $X$ be the set of active nodes of $\vec G_i$ that are
not isolated.
\item[2d.] Repeat Step 2c for local maxima, then let $X$ be the set of active
nodes of $\vec G_i$ that are not isolated. For $x\in X$, let $l_x={\it id}(x)$.
\item[2e.]
For $x\in X$, let
\[
l_x^-=\left\{
\begin{array}{ll}
l_y,&\mbox{if $y\to x$ is an edge of $\vec G_i$;}\\
l_x-1,&\mbox{if $y\to x$ is not an edge of $\vec G_i$ and $l_z>l_x$;}\\
l_x+1,&\mbox{if $y\to x$ is not an edge of $\vec G_i$ and $l_z<l_x$,}
\end{array}
\right.
\]
where $z$ is the downstream neighbor of $x$, and
\[
l_x^+=\left\{
\begin{array}{ll}
l_y,&\mbox{if $x\to y$ is an edge of $\vec G_i$;}\\
l_x+1,&\mbox{if $x\to y$ is not an edge of $\vec G_i$ and $l_z<l_x$;}\\
l_x-1,&\mbox{if $x\to y$ is not an edge of $\vec G_i$ and $l_z>l_x$,}
\end{array}
\right.
\]
where $z$ is the upstream neighbor of $x$. Now consider the binary
representations of $l_x^-$, $l_x$, and $l_x^+$, and let $A(x)$ be the set of
positive integers $p$ such that $l_x^-$ and $l_x$ have the same bit at the $p$th
position while $l_x$ and $l_x^+$ do not. Likewise, let $B(x)$ be the set of
numbers $p$ such that $l_x^-$ and $l_x$ have different bits at the $p$th
position while $l_x$ and $l_x^+$ have the same bit. Assuming that position
numbers increase from right to left in a binary representation, let $p^*(x)$ be
the greatest member of $A(x)\cup B(x)$. If $x\to y$ is an edge of $\vec G_i$
such that $p^*(x)=p^*(y)$, then combine $x$ into $y$ and make the resulting node
isolated by eliminating edges appropriately (if any node in $X$ becomes isolated
because of this, then combine that node into the newly-formed node as well). Now
let $X$ be the set of active nodes of $\vec G_i$ that are not isolated, then
repeat Steps 2c and 2d with $p^*$'s in place of ${\it id}$'s, and once again let
$X$ be the set of active nodes of $\vec G_i$ that are not isolated. If
$X\neq\emptyset$, then let $l_x=p^*(x)$ for all $x\in X$ and repeat Step 2e. If
$X=\emptyset$, then let the set of isolated nodes of $\vec G_i$ be the node set
of $\vec G_{i+1}$.
\end{enumerate}

{\em Remark.} In Step 2a, it is possible for more than one $x$ to exist for the
same $y$. In this case, every such $x$ is combined into the single resulting
node of identifier ${\it id}(y)$. Note that for no such $x$ there may exist a
node $z$ such that $z\to x$ or $y\to z$ is an edge of $\vec G_i$. This is so,
respectively, because $x$ has an inactive downstream neighbor and by Step 1c has
no active neighbors, and because $y$, being inactive, has no downstream
neighbors. As a consequence, the newly-formed node is isolated in $\vec G_i$.
At the end of Step 2a, the single downstream neighbor of every member of $X$ is
active, and therefore also a member of $X$.

{\em Remark.} In Step 2b, there may exist more than one $y\in Z(x)$ such that
$y\neq z$ and $Z(y)=\emptyset$. Every such $y$ gets combined into node $x$. At
the end of Step 2b, every member of $X$ has exactly one downstream neighbor and
at most one upstream neighbor. That is, the members of $X$ are arranged into
groups of nodes, each group having at most one node with no upstream neighbor
and exactly one node that has the same neighbor for both upstream and downstream
neighbor. Except for these two-node directed cycles, such groups of nodes may be
regarded as directed chains.

{\em Remark.} At the end of Step 2d, the members of $X$ are arranged into
directed chains of nodes whose identifiers are strictly increasing or decreasing
along the chains. Each such chain has at least two nodes, of which exactly one
has no upstream neighbor and exactly one has no downstream neighbor.

{\em Remark.} Step 2e repeatedly manipulates the node labels $l_x$ so that the
finding of minima and maxima, respectively as in Steps 2c and 2d, can once again
be used to combine nodes in $X$ into isolated nodes. Initially, node identifiers
are used for labels, but subsequently they get replaced by integers that point
into the binary representations of the labels used in the previous iteration.
As the iterations progress, these integers have an ever-dwindling range: if
$j\ge 1$ identifies an iteration within Step 2e, then the range of labels
during iteration $j$ is $0,\ldots,\log^{(j)}n$. Eventually, during a certain
iteration $j\le\log^*n$, this range becomes $\{0,1\}$ and consequently the
taking of minima and maxima as in Steps 2c and 2d is guaranteed to produce an
empty $X$. At the beginning of each iteration, the members of $X$ are arranged
into directed chains whose labels are strictly increasing or decreasing along
the chains. Each such chain has at least two nodes, of which exactly one has no
upstream neighbor and exactly one has no downstream neighbor.

Steps 1 and 2 specify the $i$th phase of Partition\_$G$ as the manipulation of
directed graphs, first to find the edges of $\vec G_i$ in Step 1, then in Step 2
to find the nodes of $\vec G_{i+1}$. Of course, the realization of such
operations on graphs requires communication among the nodes of $\vec G_i$,
which ultimately translates into communication among the nodes of $G$. However,
the assumption of a synchronous model of distributed computation makes the
communication needs of Partition\_$G$ rather straightforward to realize
\cite{p99}.

Because each node in $\vec G_i$ stands for a rooted tree in $G$, Steps 1 and 2
can be regarded as being executed by the trees' roots, which in turn coordinate
the remaining nodes in their trees in carrying out the various tasks prescribed
by the algorithm. For example, Step 1a is a broadcast with feedback on tree
edges started by the root, which sends out the upper bound of $2^{i+1}-1$ on the
tree height for the $i$th phase. This is propagated by the other nodes in the
tree, which send on what they receive, if nonzero, after decrementing it by one.
The feedback is started by the leaves, which clearly never happens if at least
one leaf is not reached by the broadcast, thus signaling to the root that the
tree is oversized.

In the same vein, by simply letting every node in $G$ that belongs to the same
node $x$ in $\vec G_i$ have a record of ${\it id}(x)$, finding the potential
neighbors of $x$ in Step 1b and the directed edges incident to it in $\vec G_i$
in Step 1c are also simple procedures that function by probing the connections
of $x$ in $G$. Whenever an edge is deployed between two nodes in $\vec G_i$, a
corresponding edge in $G$, referred to as the {\em preferred edge\/} between
those two nodes, can also be easily identified and recorded for later use.

All the remaining actions in Partition\_$G$ can be realized via communication
between the roots of two trees whose nodes in $\vec G_i$ are connected by an
edge. Whenever a message needs to be sent, it can be routed on tree edges only,
except to move from one tree to the other, at which time it must go through the
preferred edge between the two trees. This is, for example, the basis for
realizing the combination of a node into another: combining a node $x$ into a
node $y$ that is connected to it by an edge in $\vec G_i$ involves making the
preferred edge between them an edge of the new tree and then propagating through
$x$'s tree the information that a new root exists and has identifier
${\it id}(y)$.

\section{Correctness and complexity}\label{corr}

Most of our correctness and complexity arguments hinge on how well Step 2e
succeeds in breaking directed chains of nodes in $\vec G_i$ as needed. It is
to the properties of Step 2e that we turn first.

\begin{lem}\label{lemma1}
Let $x\to y$ be an edge at the beginning of an iteration of Step 2e of
Partition\_$G$. The following holds:
\begin{enumerate}
\item[(i)] If $p^*(x)\in A(x)$ and $p^*(y)\in A(y)$, then $p^*(x)\neq p^*(y)$;
\item[(ii)] If $p^*(x)\in B(x)$ and $p^*(y)\in B(y)$, then $p^*(x)\neq p^*(y)$;
\item[(iii)] If $p^*(x)\in B(x)$ and $p^*(y)\in A(y)$, then $p^*(x)\neq p^*(y)$.
\end{enumerate}
\end{lem}

{\em Proof:\/}
By Steps 2a through 2e, edge $x\to y$ is in a chain of nodes whose labels are
either strictly increasing or strictly decreasing along the chain. Suppose the
former case first. Then $l_x^-<l_x<l_y<l_y^+$.

If $p^*(x)\in A(x)$, then $l_x^-$ and $l_x$ have the same bit at position
$p^*(x)$ while $l_x$ and $l_y$ have different bits at that same position. If
$p^*(y)\in A(y)$, then $l_x$ and $l_y$ have the same bit at position $p^*(y)$
while $l_y$ and $l_y^+$ have different bits at that same position. So $p^*(x)$
and $p^*(y)$ cannot be the same position, thus proving (i).

If $p^*(x)\in B(x)$, then $l_x^-$ and $l_x$ have different bits at position
$p^*(x)$ while $l_x$ and $l_y$ have the same bit at that same position. If
$p^*(y)\in B(y)$, then $l_x$ and $l_y$ have different bits at position $p^*(y)$
while $l_y$ and $l_y^+$ have the same bit at that same position. So $p^*(x)$
and $p^*(y)$ cannot be the same position, which proves (ii).

We now prove (iii). If $p^*(x)\in B(x)$, then $l_x^-$ and $l_x$ have different
bits at position $p^*(x)$ while $l_x$ and $l_y$ have the same bit at that same
position. Suppose these bits are $100$, respectively for $l_x^-$, $l_x$, and
$l_y$. By definition of $p^*(x)$, at all other positions to the left of $p^*(x)$
in the binary representations of $l_x^-$, $l_x$, $l_y$ (that is, positions
corresponding to higher powers of two) we must have the same bit for all three
labels or bits that differ from $l_x^-$ to $l_x$ and also from $l_x$ to $l_y$.
In other words, the only possibilities are $000$, $111$, $010$, and $101$. But
these possibilities have all the same bit for $l_x^-$ and $l_y$, which
contradicts the fact that $l_x^-<l_y$. Then the bits for $l_x^-$, $l_x$, and
$l_y$ at position $p^*(x)$ must be $011$.

If $p^*(x)=p^*(y)$, then the bits of $l_x$ and $l_y$ are both $1$ at position
$p^*(y)$, which is in agreement with the definition of $A(y)$. By this same
definition, at position $p^*(y)$ the bit of $l_y^+$ must be $0$. To the left of
$p^*(y)$, the possibilities for $l_x$, $l_y$, $l_y^+$ are $000$, $111$, $010$,
and $101$, again following the definition of $p^*(y)$. This implies the same bit
for $l_x$ and $l_y^+$ at all those positions, which like before contradicts the
fact that $l_x<l_y^+$. So $p^*(x)\neq p^*(y)$.

If $x\to y$ is in a chain of nodes whose labels are strictly decreasing along
the chain, then $l_x^->l_x>l_y>l_y^+$. In this case, the arguments that prove
(i) and (ii) remain unchanged, while in the proof of (iii) it suffices to
complement every bit in the triples we displayed (this leads to contradictions
of $l_x^->l_y$ and $l_x>l_y^+$).
$\Box$

\begin{lem}\label{lemma2}
Let $x\to y$ be an edge at the beginning of an iteration of Step 2e of
Partition\_$G$. If $p^*(x)=p^*(y)$, then $p^*(x)\in A(x)$ and $p^*(y)\in B(y)$.
\end{lem}

{\em Proof:\/}
If $p^*(x)\in B(x)$ with either $p^*(y)\in B(y)$ or $p^*(y)\in A(y)$, then by
Lemma \ref{lemma1}, parts (ii) and (iii), $p^*(x)\neq p^*(y)$. If
$p^*(x)\in A(x)$ and $p^*(y)\in A(y)$, then by Lemma \ref{lemma1}, part (i),
$p^*(x)\neq p^*(y)$. Thence the lemma.
$\Box$

\begin{lem}\label{lemma3}
Let $x\to y\to z$ be part of a chain at the beginning of an iteration of Step 2e
of Partition\_$G$. If $p^*(x)=p^*(y)$, then $p^*(y)\neq p^*(z)$.
\end{lem}

{\em Proof:\/}
By Lemma \ref{lemma2}, $p^*(x)\in A(x)$ and $p^*(y)\in B(y)$. By Lemma
\ref{lemma1}, parts (ii) and (iii), $p^*(y)\neq p^*(z)$.
$\Box$

Now let $I$ be the last phase of Partition\_$G$ in which premature termination
in Step 1b does not occur. Then $1\le I\le\left\lceil\log(k+1)\right\rceil-1$
and we have the following.

\begin{lem}\label{lemma4}
For $i=0,\ldots,I$, every node of $\vec G_{i+1}$ that is not an inactive node of
$\vec G_i$ is formed by the combination of at least two nodes of $\vec G_i$.
\end{lem}

{\em Proof:\/}
By Step 2e, the node set of $\vec G_{i+1}$ is the set of isolated nodes in
$\vec G_i$ at the end of Step 2. The lemma follows easily from the fact that
every isolated node produced during Step 2 (that is, isolated nodes that are not
inactive during phase $i$) result from the combination of at least two nodes of
$\vec G_i$.
$\Box$

We are now in position to demonstrate that Partition\_$G$ does indeed achieve
its goals.

\begin{thm}\label{theorem5}
For $i=0,\ldots,I+1$, the nodes of $\vec G_i$ form a rooted spanning forest of
$G$. Each tree in this forest has at least $2^i$ nodes and $O(2^i)$ height.
\end{thm}

{\em Proof:\/}
The theorem holds trivially for $i=0$, and we prove it inductively for $i+1$
with $0\le i\le I$. The induction hypothesis is that the nodes of $\vec G_i$
form a rooted spanning forest of $G$, each of whose trees having at least $2^i$
nodes and $O(2^i)$ height.

In order to show that the nodes of $\vec G_{i+1}$ do indeed form a rooted
spanning forest of $G$, by the induction hypothesis it suffices to argue that
the set $X$ is empty at the end of phase $i$. The reason for this is that it is
the set of isolated nodes at the end of phase $i$ that we take to be the node
set of $\vec G_{i+1}$, and that $X=\emptyset$ indicates that every node in $G$
is part of the tree represented by some isolated node. But this follows directly
from the fact that the range of labels during Step 2e decreases steadily as the
iterations progress. Eventually, this range becomes such that every label is
either $0$ or $1$, at which time the finding of minima and maxima makes $X$
empty.

Having shown this, we consider the number of nodes and height of each of the
trees in the node set of $\vec G_{i+1}$. A node of $\vec G_{i+1}$ is either
an inactive node of $\vec G_i$ or results, by Lemma \ref{lemma4}, from the
combination of at least two nodes of $\vec G_i$. In the former case, by Step 1a
the node corresponds to a rooted tree in $G$ with at least $2^{i+1}$ nodes. In
the latter case, by the induction hypothesis, it corresponds to a rooted tree in
$G$ with at least $q2^i$ nodes for $q\ge 2$, that is, at least $2^{i+1}$ nodes.

As for the height, we reason similarly. If a node in $\vec G_{i+1}$ is an
inactive node of $\vec G_i$, then its height is, by the induction hypothesis,
$O(2^i)$, which in turn is $O(2^{i+1})$. If it is a combination of at least
two nodes of $\vec G_i$, then either this combination takes place in one of
Steps 2a through 2d or in Step 2e. In the former case, the combination is
either performed over a single edge (Steps 2a and 2b), or it is performed over
a chain of at most six edges (first in Step 2c, then in Step 2d). In either
case, the induction hypothesis leads to a height of $q2^i$ for $q$ a constant,
which is $O(2^{i+1})$. The case of Step 2e is entirely analogous, since by
Lemma \ref{lemma3} it is either performed over a single edge, or else by the
taking of minima and maxima, as in Steps 2c and 2d.
$\Box$

\begin{cor}\label{corollary6}
The nodes of $\vec G_{I+1}$ form a rooted spanning forest of $G$, and in this
forest each tree has at least $k+1$ nodes and $O(k)$ height.
\end{cor}

{\em Proof:\/}
If Partition\_$G$ terminates in Step 1b of some phase, then
$I<\left\lceil\log(k+1)\right\rceil-1$ and the corollary holds, because $\vec G$
has in this case one single node encompassing all the $n\ge k+1$ nodes of $G$,
and furthermore the height of the tree that corresponds to this single node is
by Theorem \ref{theorem5} $O(2^{I+1})$, which is $O(k)$. If Partition\_$G$ runs
through all the phases, then $I=\left\lceil\log(k+1)\right\rceil-1$ and the
corollary follows directly from Theorem \ref{theorem5} with
$i=I+1=\left\lceil\log(k+1)\right\rceil$.
$\Box$

We now finalize the section by discussing the time and number of messages
needed by Partition\_$G$ and by Algorithm PB as a whole.

\begin{thm}\label{theorem7}
Partition\_$G$ requires $O(k\log^*n)$ time and $O(m\log k+n\log k\log^*n)$
messages to complete.
\end{thm}

{\em Proof:\/}
During the $i$th phase, $i=0,\ldots,\left\lceil\log(k+1)\right\rceil-1$, each of
Steps 1a through 2d requires a constant number of communication ``rounds,'' each
in turn requiring a number of time units proportional to the height of a rooted
tree in that phase, which by Theorem \ref{theorem5} is $O(2^i)$. The same holds
for each of the iterations of Step 2e, of which there are at most $\log^*n$.
Then the time required for Partition\_$G$ to complete grows with
\begin{eqnarray*}
\sum_{i=0}^{\lceil\log(k+1)\rceil-1}2^i\log^*n
&\le&\sum_{i=0}^{\lceil\log(k+1)\rceil-1}\frac{2^{\log(k+1)}}{2^i}\log^*n\\
&=&(k+1)\log^*n\sum_{i=0}^{\lceil\log(k+1)\rceil-1}\frac{1}{2^i}\\
&<&2(k+1)\log^*n,
\end{eqnarray*}
so Partition\_$G$ requires $O(k\log^*n)$ time.

The number of messages that Partition\_$G$ requires can be estimated likewise
for phase $i$, as follows. The number of messages sent during Step 1 is
dominated by Step 1b to determine the potential neighbors in $G$ of a node in
$\vec G_i$, which requires $O(m)$ messages. Steps 2a and 2b require $O(n)$
messages, which is the total number of tree edges, because several nodes may be
combined into the same node during those steps. However, each of the
$O(\log^*n)$ communication ``rounds'' in Steps 2c through 2e is more economical,
because the chain structure of $\vec G_i$ in those steps allows communication to
take place along single paths from the trees' roots, and so the number of
messages flowing in each rooted tree is proportional to its height, which during
phase $i$ is $O(2^i)$ by Theorem \ref{theorem5}. Also by Theorem \ref{theorem5},
each rooted tree in $\vec G_i$ has at least $2^i$ nodes, so there are at most
$n/2^i$ rooted trees in $\vec G_i$. It follows that the number of messages
required by Partition\_$G$ for completion grows with
\[
\sum_{i=0}^{\lceil\log(k+1)\rceil-1}m+\frac{n}{2^i}2^i\log^*n
=m\left\lceil\log(k+1)\right\rceil+n\log^*n\left\lceil\log(k+1)\right\rceil,
\]
so Partition\_$G$ requires $O(m\log k+n\log k\log^*n)$ messages.
$\Box$

\begin{cor}\label{corollary8}
Algorithm PB requires $O(k\log^*n)$ time and $O(m\log k+n\log k\log^*n)$
messages to complete.
\end{cor}

{\em Proof:\/}
Immediate from Theorem \ref{theorem7}, considering that the algorithm's second
stage requires $O(k)$ time and $O(n)$ messages.
$\Box$

\section{Concluding remarks}\label{concl}

We have considered the problem of finding a $k$-dominating set with no more than
$\left\lfloor n/(k+1)\right\rfloor$ nodes in $G$, and given a new synchronous
distributed algorithm to solve it in $O(k\log^*n)$ time while requiring
$O(m\log k+n\log k\log^*n)$ messages. Our algorithm follows the same overall
strategy of \cite{kp98}, according to which first a rooted spanning forest is
found in $G$ with certain characteristics, and then the desired $k$-dominating
set on that forest.

Our algorithm introduces a new approach to finding the rooted spanning forest,
which we think is conceptually simpler than the one of \cite{kp98}, and shares
with the algorithm of \cite{kp98} the additional computation that is required to
find the $k$-dominating set. In both algorithms, the overall complexity is
dominated by the forest-finding stage. Both have the same time complexity, but
ours has better message complexity.

\ack

The authors acknowledge partial support from CNPq, CAPES, the PRONEX initiative
of Brazil's MCT under contract 41.96.0857.00, and a FAPERJ BBP grant.


\begin{thebibliography}{10}
\expandafter\ifx\csname url\endcsname\relax
  \def\url#1{\texttt{#1}}\fi
\expandafter\ifx\csname urlprefix\endcsname\relax\def\urlprefix{URL }\fi

\bibitem{wz01}
R.~Wittmann, M.~Zitterbart, Multicast Communication: Protocols and
  Applications, Morgan Kaufmann Publishers, San Francisco, CA, 2001.

\bibitem{bkp93}
J.~Bar-Ilan, G.~Kortsarz, D.~Peleg, How to allocate network centers, Journal of
  Algorithms 15 (1993) 385--415.

\bibitem{p90}
D.~Peleg, Distributed data structures: a complexity-oriented view, in:
  Proceedings of the Fourth International Workshop on Distributed Algorithms,
  1990, pp. 71--89.

\bibitem{pu89}
D.~Peleg, E.~Upfal, A tradeoff between size and efficiency for routing tables,
  Journal of the ACM 36 (1989) 510--530.

\bibitem{gj79}
M.~R. Garey, D.~S. Johnson, Computers and Intractability: A Guide to the Theory
  of NP-Completeness, W. H. Freeman \& Co., New York, NY, 1979.

\bibitem{kp98}
S.~Kutten, D.~Peleg, Fast distributed construction of small $k$-dominating sets
  and applications, Journal of Algorithms 28 (1998) 40--66.

\bibitem{p99}
L.~D. Penso, A distributed algorithm to find $k$-dominating sets in graphs,
  Master's thesis, Federal University of Rio de Janeiro, in Portuguese
  (December 1999).

\bibitem{b96}
V.~C. Barbosa, An Introduction to Distributed Algorithms, The MIT Press,
  Cambridge, MA, 1996.

\bibitem{gkp98}
J.~A. Garay, S.~Kutten, D.~Peleg, A sub-linear time distributed algorithm for
  minimum-weight spanning trees, SIAM Journal on Computing 27 (1998) 302--316.

\bibitem{gps87}
A.~V. Goldberg, S.~A. Plotkin, G.~E. Shannon, Parallel symmetry-breaking in
  sparse graphs, in: Proceedings of the Nineteenth Annual ACM Symposium on
  Theory of Computing, 1987, pp. 315--323.

\bibitem{ps92}
A.~Panconesi, A.~Srinivasan, Improved distributed algorithms for coloring and
  network decomposition problems, in: Proceedings of the Twenty-Fourth Annual
  ACM Symposium on Theory of Computing, 1992, pp. 581--592.

\end{thebibliography}

\end{document}